\documentclass[usenatbib]{basi}
\usepackage[T1]{fontenc}
\usepackage[british]{babel}
\usepackage[varg]{txfonts}
\usepackage{rotating}
\usepackage{dcolumn}
\def\mnras{MNRAS}
\def\apj{ApJ}
\begin{document}
\title[Speca and RAD@home]{New results on the exotic galaxy `Speca' and discovering many more Specas with RAD@home network 
}
\author[Hota et~al.]%
{\parbox{\textwidth}{       
Ananda Hota$^{1,2}$\thanks{email: \texttt{hotaananda@gmail.com, radathomeindia@gmail.com}}, Judith H. Croston$^{3}$, Youichi Ohyama$^{4}$, C. S. Stalin$^{5}$, 
Martin J. Hardcastle$^6$, Chiranjib Konar$^{4}$, R.P. Aravind$^{2}$, Sheena M. Agarwal$^{2}$, Sai Arun Dharmik Bhoga$^{2}$, 
Pratik A. Dabhade$^{2}$, Amit A. Kamble$^{2}$, Pradeepta K. Mohanty$^{2}$, Alok Mukherjee$^{2}$, Akansha V. Pandey$^{2}$, 
Alakananda Patra$^{2}$, Renuka Pechetti$^{2}$, Shrishail S. Raut$^{2}$,  V. Sushma$^{2}$, Sravani Vaddi$^{2}$, Nishchhal Verma$^{2}$}
\vspace{0.4cm}
\\
\parbox{\textwidth}{
       $^1$UM-DAE Centre for Excellence in Basic Sciences (CBS), Vidyanagari, Mumbai-98, India\\
       $^2$RAD@home Astronomy Collaboratory, India\\
       $^3$School of Physics and Astronomy, University of Southampton, Southampton, SO17 1BJ, UK\\
       $^4$Institute of Astronomy and Astrophysics, Academia Sinica, Taipei-106, Taiwan\\
       $^5$Indian Institute of Astrophysics, Bangalore 560 034, India\\
$^6$ School of Physics, Astronomy and Mathematics, University of Hertfordshire, Hatfield, UK
}}
\pubyear{2014}
\volume{00}
\pagerange{\pageref{firstpage}--\pageref{lastpage}}
\date{Received --- ; accepted ---}
\maketitle
\label{firstpage} 
\begin{abstract}
We present the first report on an innovative new project named "RAD@home", a citizen-science
research collaboratory built on free web-services like Facebook, Google, Skype, NASA Skyview, NED, TGSS etc..
This is the first of its kind in India, a zero-funded, zero-infrastructure, human-resource network 
to educate and directly involve in research, hundreds of science-educated under-graduate population of India, 
irrespective of their official employment and home-location with in the country. Professional international
collaborators are involved in follow up observation and publication of the objects discovered by 
the collaboratory. We present here ten newly found candidate episodic radio galaxies, already proposed to GMRT, 
and ten more interesting cases which includes, bent-lobe radio galaxies located in new Mpc-scale filaments, 
likely tracing cosmological cluster accretion from the cosmic web. Two new Speca-like rare  
spiral-host large radio galaxies have also been been reported here. Early analyses from our follow up observations 
with the Subaru and XMM-Newton telescopes have revealed that Speca is likely a new entry to the cluster and is a 
fast rotating, extremely massive, star forming disk galaxy. Speca-like massive galaxies with giant radio lobes, are 
possibly remnants of luminous quasars in the early Universe or of first supermassive black holes with in first masssve galaxies. 
As discoveries of Speca-like galaxies did not require new data from big telescopes, but free archival radio-optical data, 
these early results demonstrate the discovery potential of RAD@home and how it can help resource-rich professionals, as well as 
demonstrate a model of academic-growth for resource-poor people in the underdeveloped regions via Internet.
\end{abstract}
\section{Update on Speca from XMM-Newton and Subaru}
Following discovery of the spiral-host episodic giant radio galaxy `Speca' \citep{hota11}, it has been observed with the Subaru telescope in 
long-slit spectroscopy mode and with the XMM-Newton for investigating the kinematics and cluster-environment it belongs to, respectively. 
H$\alpha$ and [N II] emission lines seen from the major axis of the galaxy revealed that it is an extremely fast-rotating (maximum speed 
$\sim$370 km s$^{-1}$ near the flat regime), star forming disk galaxy. Though it is the brightest member of the identified cluster, the point-source 
subtracted diffuse X-ray emission is offset from Speca and co-spatial with the early-type members of the cluster. With the discovery of a second such 
low-z, fast-rotating, spiral-host, episodic giant radio galaxy (Bagchi et al. 2014 presented in this conference), it is clear that they may be rare 
but it forms a different class of active galaxies. 
Presence of large number of starbursting, molecular gas-rich (with double-horn profile) radio galaxies at high-z and 
presence of luminous quasars with billion solar mass supermassive black holes of early Universe suggest their remnants to be found at the low-z Universe. 
We hypothecise that Speca-like spiral-host, massive, giant radio galaxies are the possible remnants of these first massive galaxies with supermassive 
black holes, who have avoided major mergers (with big spirals and into clusters) and have become radio loud only recently. So, finding more Speca-like 
galaxies is like spotting `living fossils' and it is important to investigate their nature, origin with more discoveries for better understanding of 
black hole, galaxy and cluster evolution. \\
\section{ Launching of and Discoveries from RAD@home} 
Three pieces of international news, namely {\bf 1.} Discovery of Speca (\citep{hota11}; NRAO (NSF) \& NCRA-TIFR press release) 
{\bf 2.} AGN-jet feedback caught in the act or Cosmic leaf-blower galaxy NGC3801 (\citep{hota12}; NASA-JPL-CalTech news release) 
and {\bf 3.} Discovery of a blue supergiant star (\citep{hota13}; Subaru telescope (NAOJ) press release), created nearly a million Google hits, in total, 
and nearly one thousand Orkut and Facebook shares (648 shares of the article in the TIME magazine website on NGC3801). 
There is a huge potential in connecting with numerous astronomy readers, 
spread all across profession and location, which can help the astronomers in solving the "big data" problem, as demonstrated in the 
"Galaxy-zoo" citizen-science project. The lead author realised that the reverse is also possible. 
The network can help academic-growth of deserving students and teachers in the remote regions and quenching intellectual thirst of 
astronomy-enthusiasts(amateur-astronomers), who are unemployed (e.g. educated housewives) or were unable to pursue a PhD career in astronomy 
due to socio-economic or geo-political problems. Taking advantage of the third news, a Facebook group was created on 15th April 2013 
({\it https://facebook.com/groups/RADathome/}). In ten months it has attracted over six hundred Indians with 
undergraduate education in science or engineering (with a tag line: Any BSc/BE Can Do Research -- ABCDR). 
They have been trained to analyse UV-optical-IR-radio multi-wavelength RGB-images using NASA Skyview 
({\it http://skyview.gsfc.nasa.gov/}) and use of TIFR GMRT Sky Survey 
(TGSS) images using SAO ds9. Images of well-known galaxies as well as images of new galaxies (without disclosing the co-ordinate) have been shared 
and discussed in the group. Services of Google (gmail, drive, hangout (screen-share), youtube) and Skype were used by the principal investigator 
(PI) of RAD@home, the lead author, to train and interact with the leading members for assessing authenticity and importance of the new-found objects. \\
{\bf Candidate Episodic Radio Galaxies:} Three patterns namely 1. disk-like host galaxy 2. episodic radio lobes 3. relic radio lobes in a dynamic
intra-cluster medium, were explained to members hunting for another Speca. The following objects were spotted for possible episodic nature. 
Names of these episodic radio galaxies (ERG) and their J2000.0 coordinates, as proposed in the GMRT proposal, are RAD-1 (03 16 03.0, -26 57 55: 
double double radio galaxy (DDRG)), RAD-2 (05 12 13.1, +01 31 49: DDRG), RAD-3 (08 47 23.9, -27 16 42: X-shaped ERG), RAD-4 (10 42 05.8 -23 38 06: 
DDRG), RAD-5 (10 46 32.4, -01 13 37: X-shaped ERG), RAD-6 (11 02 29.4, -33 42 34: DDRG), RAD-7 (11 30 41.0, -03 48 07: S-shaped ERG), 
RAD-8 (11 57 36.0, -03 14 41: DDRG), RAD-9 (12 32 46.0, -00 55 25: DDRG), RAD-10 (12 37 45.9, -01 14 17: DDRG). RAD-7 (Fig. 1) is an intriguing case where 
the possible host is extremely elongated, possibly a disk but asymmetrically located to the bright lobes seen to VLA FIRST image. Beyond these 
two lobes faint and S-shaped radio emission extends like the case of Hydra-A. It is a case of precession, galaxy merger or a distant cluster 
is unclear with the available data. Its similarity with Speca and unknown redshift require radio-optical follow ups which are is in progress.\\
\begin{figure}
\centerline{\includegraphics[width=9cm]{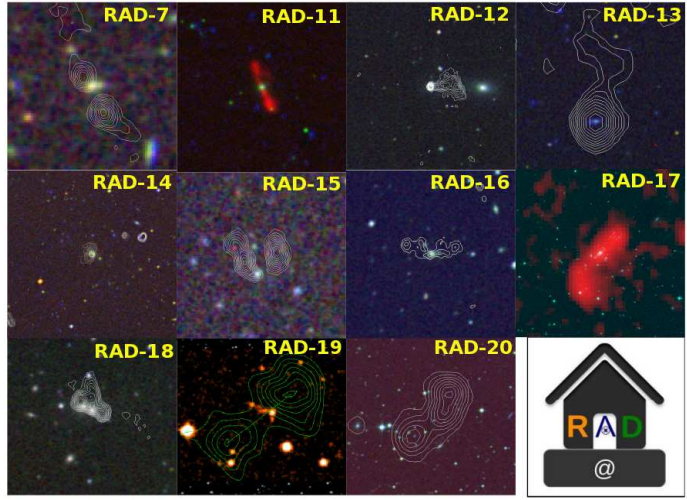}}
\caption{Multi-wavelength (UV-Opt-IR-radio) images of RAD-7 (11 30 41.0, -03 48 07), RAD-11 (00 21 07.6, -00 55 31), RAD-12 (00 43 00.6 -09 13 46),
RAD-13 (00 53 09.9, -10 11 11), RAD-14 (01 12 55.1, -09 50 41), RAD-15 (11 27 15.0, -00 23 44), RAD-16 (11 45 26.0, -02 23 33),
RAD-17 (13 02 03.6, -00 50 12), RAD-18 (13 30 10.3 -02 06 18), RAD-19 (21 09 59.5, -26 36 24), RAD-20 (21 25 55.2 -33 13 18),
made using Skyview, ds9 and TGSS. Official logo of RAD@home is also added to it}
\end{figure}
{\bf Interesting cases of black hole-galaxy and cluster evolution:} While searching for another Speca the following interesting cases were found (Fig. 1). 
{\bf RAD-12} is an extreme case where the radio lobe is likely hitting inter-stellar medium of another nearby galaxy (at same redshift) on its 
west and has brightened up significantly compared to the eastern counterpart, if any. On the other hand in {\bf RAD-17} the eastern lobe, far away from the
host galaxy, makes an almost 90$^{\circ}$ turn and no optical galaxy is seen to have deflected the radio plasma like that, 
an extreme case of jet-IGM/ICM interaction. 
{\bf RAD-16} is a bent-lobe radio galaxy, which is sitting in the middle of a nearly 2 Mpc wide filament of galaxies. Only one such clear case is 
known in the literature and further study is in progress. {\bf RAD-14} is also a similar case of bent-lobe radio galaxy where the bending has 
likely been due to ram pressure of the cosmological accretion flow towards east where accumulation of a large number of galaxies 
with similar redshift is seen, suggesting a nearby cluster or large-scale structure. {\bf RAD-11} was noted as a case of lateral displacement of 
radio plasma (towards the south-west) where presence of numerous small UV-bright blue galaxies were located in the surrounding, a case to 
look for possible accretion on to near by cluster through a filament. {\bf RAD-13} is a case of infall of the galaxy to possibly a nearby cluster 
where the optical host as well as tail in the radio emission both suggest strong effect of ram pressure stripping. 
{\bf RAD-19} is a case of relic-lobe or dead-lobe radio galaxy where the host galaxy is seen to be undergoing a tidal interaction with a companion 
and strangely the tidal stellar bridge is relatively bright in IR bands than blue optical/UV bands. 
{\bf RAD-18} is again a rare opportunity where, two early-type galaxies are in the process of merging and one of them is radio loud. 
A case of dry-merger to have counter-intuitively supplied gas and has possibly triggered the AGN-activity. 
{\bf RAD-15} is case of two slightly curved, "()" shaped radio sources, symmetric and co-spatial with a galaxy cluster. 
However the cluster is larger than the curved radio sources and the peaks of the radio emission does not show any optical counterparts. 
Its similarity with the relic-radio sources is noted but could also be two high-z FR I radio galaxies. 
{\bf RAD-20} again a rare-case and a candidate for Speca-like galaxy, where the host galaxy is bright in near-UV emission and the IR-image shows the host to be 
extremely elongated, either case of a disk host or merger. The large radio lobes are also bent suggesting infall into a cluster or in a merging groups. \\
{\bf Acknowledgements} We are grateful to Profs Govind Swarup, S.M. Chitre and S. Ananthakrishnan for encouragement. We acknowledge UM-DAE CBS 
for supporting the academic visits of some of the members of the collaboratory. 

\label{lastpage}

\begin{thebibliography}{}
{\small
\bibitem[\protect\citeauthoryear{{Hota} et~al.}{{Hota} et~al.}{2011}]{hota11}
{Hota} Ananda, {Sirothia} S.K., {Ohyama} Youichi; {Konar} C., {Kim} Suk, {Rey} Soo-Chang, {Saikia} D.J., {Croston} J.H., {Matsushita} Satoki 2011, \mnras, 417L, 36
\bibitem[\protect\citeauthoryear{{Hota} et~al.}{{Hota} et~al.}{2012}]{hota12}
{Hota} Ananda, {Rey} Soo-Chang, {Kang} Yongbeom, {Kim} Suk, {Matsushita} Satoki, {Chung} Jiwon, 2012, \mnras, 422L, 38
\bibitem[\protect\citeauthoryear{Ohyama \& Hota}{ Ohyama \& Hota}{2013}]{hota13}
{Ohyama} Youichi \& {Hota} Ananda, 2013, \apj, 767L, 29
}
\end{thebibliography}
\end{document}